\documentclass[sigconf]{acmart}
\def\BibTeX{{\rm B\kern-.05em{\sc i\kern-.025em b}\kern-.08emT\kern-.1667em\lower.7ex\hbox{E}\kern-.125emX}}
    
\usepackage{algorithm,algorithmic}
\usepackage{subfig}

\begin{document}

%
\title[Parallel transport TDDFT with hybrid functional on Summit]{Parallel Transport Time-Dependent Density Functional Theory Calculations with Hybrid Functional on Summit}

%
\author{Weile Jia}
\email{jiaweile@berkeley.edu}
\affiliation{%
  \institution{University of California, Berkeley}
  \city{Berkeley}
  \state{California}
  \postcode{94720}
}

 \author{Lin-Wang Wang}
\email{lwwang@lbl.gov}
\affiliation{%
  \institution{Lawrence Berkeley National Laboratory}
  \city{Berkeley}
  \state{California}
  \postcode{94720}
}

\author{Lin Lin}
\authornote{Corresponding author}
\email{linlin@math.berkeley.edu}
\affiliation{%
  \institution{University of California, Berkeley \\ Lawrence Berkeley National Laboratory}
  \city{Berkeley}
  \state{California}
  \postcode{94720}
}

%
\newcommand{\Or}{\mathcal{O}}
\newcommand{\jump}[1]{\big[\hspace{-0.7mm} \big[ #1 \big]
  \hspace{-0.7mm} \big]}
\newcommand{\mean}[1] {\big\{ \hspace{-0.7mm} \big\{ #1 \big\}
  \hspace{-0.7mm} \big\}}
\newcommand{\abs}[1]{\left\lvert#1\right\rvert}
\newcommand{\norm}[1]{\left\lVert#1\right\rVert}
\newcommand{\average}[1]{\left\langle#1\right\rangle}
\newcommand{\bra}[1]{\langle#1\rvert}
\newcommand{\ket}[1]{\lvert#1\rangle}
\newcommand{\mc}[1]{\mathcal{#1}}
\newcommand{\DG}{\mathrm{DG}}
\newcommand{\ud}{\,\mathrm{d}}

\newcommand{\bd}[1]{\boldsymbol{#1}}
\newcommand{\wt}[1]{\widetilde{#1}}
\newcommand{\wh}[1]{\widehat{#1}}
\newcommand{\wb}[1]{\overline{#1}}

\newcommand{\onlinecite}[1]{\citenum{#1}}
\newcommand{\bvec}[1]{\mathbf{#1}}
\newcommand{\vr}{\bvec{r}}
\newcommand{\vR}{\bvec{R}}
\newcommand{\I}{\mathrm{i}} 

\newcommand{\vace}{\widetilde{V}_{\mathrm{X}}}
\newcommand{\kace}{K^{\mathrm{ACE}}}
\newcommand{\ext}{\mathrm{ext}}
\newcommand{\Hxc}{\mathrm{Hxc}}
\newcommand{\X}{\mathrm{X}}
\newcommand{\xc}{\mathrm{xc}}
\newcommand{\eff}{\mathrm{eff}}

\newcommand{\LL}[1]{{\color{blue}~\textsf{[LL: #1]}}}
\newcommand{\WL}[1]{\textcolor{cyan}{[WL:#1]}}

\newcommand{\REV}[1]{{#1}}
\newcommand{\angstrom}{\mbox{\normalfont\AA}~}

%
\begin{abstract}

Real-time time-dependent density functional theory (rt-TDDFT) with hybrid exchange-correlation functional has wide-ranging applications in chemistry and material science simulations. However, it can be thousands of times more expensive than a conventional ground state DFT simulation, hence is limited to small systems. In this paper, we accelerate hybrid functional rt-TDDFT calculations using the parallel transport gauge formalism, and the GPU implementation on Summit. Our implementation can efficiently scale to 786 GPUs for a large system with 1536 silicon atoms, and the wall clock time is only 1.5 hours per femtosecond. This unprecedented speed enables the simulation of large systems with more than 1000 atoms using rt-TDDFT and hybrid functional. 
\end{abstract}

%
%
%

%

%
\maketitle

\section{Introduction}
Real-time time-dependent density functional theory(rt-TDDFT) is one of the newest trends in electronic structure calculations~\cite{RungeGross1984,OnidaReiningRubio2002,YabanaBertsch1996,Ullrich2011,AndradeAlberdi-RodriguezStrubbeEtAl2012}. Its popularity rises together with the recent experimental emphasis in electronic ultrafast phenomena in material science. It can be used to study ion collision, light absorption spectrum, laser-induced demagnetization and phase change, charge transfer, excited carrier dynamics, and chemical reactions. 
Rt-TDDFT will be even more powerful when it can be combined with the hybrid exchange-correlation functional~\cite{Becke1993,PerdewErnzerhofBurke1996} within the density functional theory (DFT). The hybrid functional mixes a fraction of the explicit Fock exchange integral with the semi-local exchange correlation functionals. It can be used to accurately describe the band gaps for a wide range of materials, especially with the recently developed range-separated hybrid functional~\cite{HeydScuseriaErnzerhof2003,HeydScuseriaErnzerhof2006}. As a result, the rt-TDDFT + hybrid functional approach can be extremely powerful to describe, for example, the exciton excitation and charge transfer processes. Unfortunately, both rt-TDDFT and hybrid functional is extremely computational expensive compared to conventional ground state DFT calculations with semi-local functionals. In practice, rt-TDDFT can be hundreds of times slower than the conventional ground state molecular dynamics simulations due to the need for using small time step, hybrid functional can also be tens of times slower than the semi-local exchange correlation functional due to the evaluation of the Fock exchange term. As a result, 
planewave-based rt-TDDFT + hybrid functional simulations are rarely found in the literature, even for small systems, not to mention large ones with a thousand atoms. However, for many problems, e.g., for excited state charge transfer, large system simulation is essential. Besides, planewave basis is also important due to its flexibility, especially for excited states.

Fortunately, the situation has been improved due to both new algorithm developments and the emergence of new computer platforms like Summit (located in Oak Ridge National Laboratory, listed as No.1 supercomputer in the Top500 list in November 2018).  The recently introduced parallel transport gauge formalism (PT)~\cite{JiaAnWangEtAlPT,AnLinPT} can increase the time step from the original sub-attosecond regime to around 50 attoseconds, while the Summit machine equipped with massive GPU processors has significantly increased the available computing power. Given this situation, it will be of paramount interest to test the limit of the rt-TDDFT+hybrid functional method on the Summit supercomputer. Given the heterogeneous architecture of the Summit(GPU+CPU), its latest hardware for data communication, and the large amount of total memory, the computer algorithms need to be adapted accordingly. Here, we show that, by implementing the new PT algorithm on GPUs, and by taking advantage of the large communication bandwidth and the large amount of memory on the machine, we can carry out an rt-TDDFT+hybrid functional simulation with the planewave basis set for a unprecedentedly large system with 1536 silicon atoms, with a practical time to solution of 1.5 hours per femtosecond on 768 GPUs. 

We remark that in the context of ground state hybrid functional DFT calculations, several approaches have been proposed to reduce the cost. The application of the Fock exchange operator can be accelerated with a massive number of CPUs~\cite{DucheminGygi2010,ValievBylaskaGovindEtAl2010}. The GPU acceleration has been reported in software packages such as ABINIT~\cite{GONZE2016}, BigDFT~\cite{Ratcliff_2018}, NWChem~\cite{VALIEV20101477}, Octopus~\cite{Andrade_2012}, PWmat~\cite{JIA2013, JIA2013102}, Quantum ESPRESSO~\cite{RomeroJoshua2018},  VASP~\cite{hacene2012accelerating, hutchinson2012vasp}, to name a few. When approximation of the Fock exchange operator can be tolerated, the cost can also be reduced through other algorithmic approaches such as localization~\cite{WuSelloniCar2009,DamleLinYing2015,DawsonGygi2015,CarnimeoBaroniGiannozzi2018} and density fitting techniques~\cite{HuLinYang2017}.  

Through our study, we like to address the following questions: (1) How scalable is the implementation, both in terms of strong scaling and weak scaling?  (2) How large is the speedup comparing GPU with CPU, in the sense of the absolute fastest time to solution, and in the sense of the same power consumption? (3) What is the bottleneck in the most scalable case: computation or communication? (4) Whether the memory is a bottleneck? If not, what new algorithms one can use to take advantage of the large amount of memory? Through our study, we found that, the large capacity of the Summit machine allows us to implement some unique algorithms, such as the Anderson mixing for the wavefunctions, where up to 20 copies of wavefunctions are required; also the algorithm of the Fock exchange integral evaluations (Eq.~\ref{eqn:applyVX} and Alg.~\ref{alg:Fockalg}), where a wavefunction is broadcast in an as-needed basis, and the computation can overlap with communication. The use of the new PT algorithm is essential. We show that the PT algorithm can be 20--30 times faster than the more conventional Runge-Kutta 4th order (RK4) method, and it can be efficiently implemented on supercomputers like Summit. The performance of the hardware also determines the choice of optimal algorithms. For example, recently it has been shown that in CPU machines, the adaptively compressed exchange (ACE) algorithm~\cite{Lin2016ACE} can be combined with the PT formulation to reduce the time for rt-TDDFT + hybrid functional calculations~\cite{JiaLin2019}. In this work, we find that with the GPU acceleration, the use of the PT formulation alone leads to more efficient implementation on the Summit machine. We provide detailed performance analysis, which gives insights for how to implement similar electronic structure codes in such heterogeneous platforms, and what aspects of such platforms can be improved in the future to serve similar applications.

The rest of the manuscript is organized as follows. We review the algorithm for performing 
hybrid functional rt-TDDFT calculations with the parallel transport gauge
formulation in section~\ref{sec:rt-tddft}. 
The GPU implementation is shown in section ~\ref{sec:gpu}. The setup of the test systems
and the machine configuration are presented in section ~\ref{sec:system} and
section ~\ref{sec:machine}, respectively. Then we show the numerical results
in section ~\ref{sec:result}, followed by the analysis in section ~\ref{sec:analysis}
and conclusion in section ~\ref{sec:conclusion}.

\section{Parallel transport gauge formulation of rt-TDDFT} \label{sec:rt-tddft}

Real-time time-dependent density functional theory solves the following set of time-dependent equations
\begin{equation} 
  \I \partial_{t} \Psi(t) = H(t,P(t)) \Psi(t).
  \label{eqn:tddft}
\end{equation}
Here 
$\Psi(t)=[\psi_{1}(t),\ldots,\psi_{N_{e}}(t)]$ is the collection of  electron orbitals, and $N_e$ is the number of electrons (spin degeneracy omitted). The time-dependent Hamiltonian takes the form
\begin{equation}
  H(t,P(t)) = -\frac12 \Delta_{\vr} + V_{\ext}(t) + V_{\Hxc}[P(t)] +
  V_{\X}[P(t)].
  \label{}
\end{equation}
Here $V_{\ext}(t)$ includes the contribution from the external field (such as the laser field), together with the local and nonlocal contribution from the pseudopotentials.  The Hamiltonian also depends nonlinearly on the density matrix $P(t) =
\Psi(t)\Psi^{*}(t)$.  $V_{\Hxc}$ is a local operator and characterizes the Hartree contribution and the local and the semi-local part of the exchange-correlation contribution. 
The Fock exchange operator $V_{\X}$ is an integral operator with kernel $V_{\X}[P](\vr,\vr') = -\alpha P(\vr,\vr') K(\vr-\vr')$.
Here $K(\vr-\vr')$ is the kernel for the (possibly screened) electron-electron interaction~\cite{HeydScuseriaErnzerhof2003,HeydScuseriaErnzerhof2006},
and $\alpha$ is a mixing fraction.

In an rt-TDDFT simulation, the matrix-vector multiplication of the type $H[P]\Psi$ needs to be repeatedly performed. This is particularly the case for hybrid functional calculations, where each set of multiplications $V_X[P]\Psi$ requires the following operations:
\begin{equation}
  \left(V_{X}[P]\psi_j\right)(\vr) = 
  -\sum_{i=1}^{N_{e}} \psi_{i}(\vr,t) \int
  K(\vr-\vr')\psi_{i}^{*}(\vr',t)\psi_j(\vr') \ud \vr'.
  \label{eqn:applyVX}
\end{equation}
This amounts to solving $N_{e}^2$ Poisson-like equations. Due to the convolutional structure of $K$, this can be efficiently performed using the fast Fourier transform (FFT). In practice, the application of the Fock exchange operator usually takes about $95\%$ of the total computation time.

The rt-TDDFT equation~\eqref{eqn:tddft} can be equivalently expressed using a set of unitarily transformed orbitals. Physical observables such as the density matrix are by definition gauge-invariant. This allows us to seek for the optimal gauge for numerical simulation of rt-TDDFT. Recently, such optimal gauge has been identified~\cite{JiaAnWangEtAlPT,AnLinPT}, which is defined implicitly through the following equation
\begin{equation}
  \I \partial_t \Psi = H\Psi -
  \Psi(\Psi^{*}H\Psi), \quad P(t) = \Psi(t)\Psi^{*}(t).
  \label{eqn:pt}
\end{equation}
Here $\Psi^*$ stands for the Hermitian conjugate of the matrix $\Psi$.
Compared to Eq.~\eqref{eqn:tddft}, the only difference is the extra term  $\Psi(\Psi^{*}H\Psi)$, which mixes the information from all orbitals together. The right-hand side of Eq.~\eqref{eqn:pt} is a residual type term. Its magnitude can be much smaller than $H\Psi$ on the right-hand side of  Eq.~\eqref{eqn:tddft}, and the dynamics become smoother. In fact, $\Psi(t)$ solved from Eq.~\eqref{eqn:tddft} can be viewed as the parallel transport (PT) of the initial wavefunctions to time $t$ within the range of $P(t)$, and the corresponding implicitly defined gauge is called the parallel transport gauge. It can be proved that the parallel transport gauge yields the slowest possible dynamics among all possible choices of the gauge~\cite{JiaAnWangEtAlPT}. Coupled with implicit integrators such as the Crank-Nicolson scheme, the resulting PT-CN scheme solves the following nonlinear equation at each time step
\begin{equation}\label{eqn:ptcnApp}
  \begin{split}
    &\Psi_{n+1} + \I \frac{\Delta t}{2} \left\{
    H_{n+1} 
    \Psi_{n+1} - \Psi_{n+1}\left(\Psi_{n+1}^{*}
    H_{n+1} \Psi_{n+1}\right)\right\} \\
    = &\Psi_{n} -\I \frac{\Delta t}{2} \left\{
    H_{n}
    \Psi_{n} - \Psi_{n}\left(\Psi_{n}^{*}
    H_{n} \Psi_{n}\right)\right\}.
  \end{split}
\end{equation}
Compared to explicit time integrators such as the explicit 4th
order Runge-Kutta scheme (RK4) which often requires a sub-attosecond time step, the time step allowed by PT-CN can be significantly improved to around 10--50 attoseconds. This is particularly important for reducing the number of Fock exchange operator applications in hybrid functional calculations. 

Alg.~\ref{alg:PTCN} summarizes the procedure for one step of time propagation using the PT-CN scheme. During each time step, we first evaluate the initial residual $R_n$. The right-hand side of~\eqref{eqn:ptcnApp} can be viewed as propagating the wavefunctions by half a step. It is thus denoted by $\Psi_{n+\frac12}$ and is fixed during the self-consistent field iteration. The new set of wavefunction $\Psi_{n+1}$ needs to satisfy a fixed point problem and is denoted by $\Psi_f$ during the iteration, and the residual for the fixed point problem is denoted by $R_f$. The fixed point problem is solved by the Anderson mixing method~\cite{Anderson1965}. When the residual is sufficiently small, the SCF iteration can be terminated. In practice, we find that the SCF convergence can also be monitored by the convergence of the charge density.

\begin{algorithm}
  \caption{One time propagation step with the PT-CN method.}
\begin{flushleft}
        \textbf{INPUT:}  {$\Psi_n$} \\
        \textbf{OUTPUT:} {$\Psi_{n+1}$} 
\end{flushleft}

  \begin{algorithmic}[1]
    \STATE Evaluate the initial residual $R_n = H_n\Psi_n - \Psi_n ( \Psi_n^{*} H_n \Psi_n )$.
    \STATE Evaluate $\Psi_{n+\frac12} = \Psi_n - \frac{\I \Delta t}{2} R_n$, and let $\Psi_f = \Psi_{n+\frac12}$.
    \STATE Evaluate the electron density $\rho_f$ corresponding to $\Psi_f$.
    \FOR {$j = 1, 2, \ldots$}
    \STATE Update the potential and the Hamiltonian $H_f$.
    \STATE Evaluate the fixed point residual $R_f = \Psi_f + \frac{\I \Delta t}{2}(H_f \Psi_f-\Psi_f (\Psi_fH_f \Psi_f))-\Psi_{n+\frac12}$.
    \STATE Perform Anderson mixing to update wavefunctions $\Psi_f$.
    \STATE Evaluate the electron density $\rho_f$ corresponding to $\Psi_f$. 
    \STATE If the change of the electron density is sufficiently small, exit the loop.
    \ENDFOR
    \STATE Orthogonalize {$\Psi_f$} to obtain $\Psi_{n+1}$.
  \end{algorithmic}

  \label{alg:PTCN}
\end{algorithm}



\section{Multi-GPU implementation}\label{sec:gpu}

In hybrid functional calculations with a planewave basis set, the application of the Fock exchange operator in the $H\Psi$ step often takes around $95\%$ of the total computation time with a CPU implementation. However, according to Amdahl's law, in order to achieve a desirable speedup factor, almost all steps of the calculation needs to be accelerated using GPUs. Our implementation is based on PWDFT, which uses the planewave discretization and is an independent module of the massively parallel software package DGDFT (Discontinuous Galerkin Density Functional Theory)~\cite{LinLuYingE2012,HuLinYang2015a}. In our implementation, all computationally intensive parts are performed using GPUs with either GPU-accelerated libraries or CUDA custom kernels. We also carefully overlap the MPI communication and GPU computation to take advantage of the heterogeneous architecture. 

\subsection{Hybrid parallelization scheme}

There are two main parallel distribution schemes for the wavefunctions. The first one is the column based distribution scheme (also called the band index parallelization), i.e., each column of $\Psi$ are distributed to different MPI tasks based on its band index (i.e. the column index). This data distribution scheme is highly efficient for the calculation of $H\Psi$. This is particularly the case for hybrid functional calculations since different MPI tasks are able to perform FFTs independently. The second one is the row based distribution scheme (also called the $G$-space parallelization, where $G$ is a standard notation for the index in the Fourier space). In this scheme, the data is distributed according to the partition of the Fourier coefficients. This distribution scheme facilitates the evaluation of matrix-matrix multiplications, such as the evaluation of the overlap matrix $S=\Psi^* (H\Psi)$ once $H\Psi$ is obtained. 

\begin{figure}[h]
  \begin{center}{\includegraphics[width=0.35\textwidth]{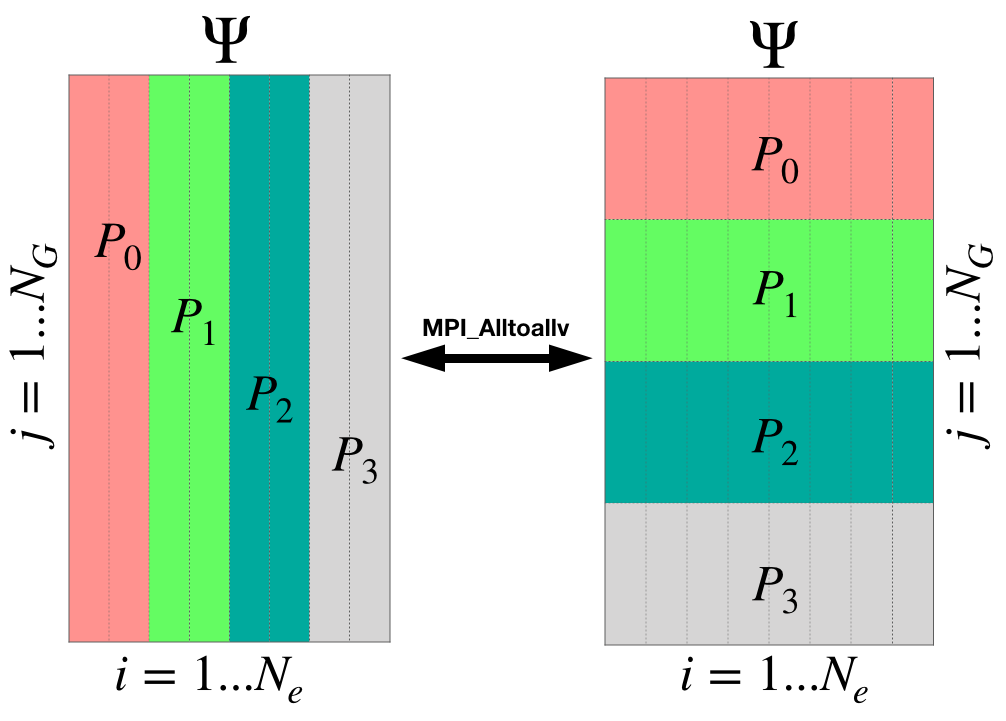}} \end{center}
  \caption{Illustration of the hybrid parallelization scheme with $N_e=8$ wavefunctions on 4 MPI tasks. The number of planewaves $N_G$ is usually on the order of $10^{3}\sim 10^{6}$.    Thus the maximum number of processes is limited by the $N_e$.}
  \label{fig:hybrid_para}
\end{figure}

In PWDFT, the wavefunctions $\Psi$ are mostly distributed in the band index parallelization to evaluate the $H\Psi$ efficiently. The conversion from band index parallelization to $G$-space parallelization is performed via \textsf{MPI\_Alltoallv} for matrix-matrix multiplication operations such as the evaluation of the overlap matrix, as shown in Fig.~\ref{fig:hybrid_para}. Note that after the evaluation in the $G$-space, the wavefunctions are converted back to the band index parallelization format via \textsf{MPI\_Alltoallv}. In this fashion, both $H\Psi$ and matrix-matrix multiplication operations can be evaluated efficiently. 
In the GPU implementation,  the hybrid parallelization scheme plays an even more important role. This is because the band index parallelization allows us to use the CUFFT  library for the $H\Psi$ calculation. However, the band index parallelization cannot scale to a large number of processors for matrix-matrix multiplications either on CPUs or GPUs, and we therefore need to convert to the $G$-space parallelization scheme. The hybrid parallelization scheme for the GPU implementation has been used in~\cite{WangEtAl2011} for ground state electronic structure calculations. 
In the discussion above, we assume only the $\Gamma$ point is considered for sampling the Brillouin zone. For solid state systems with $k$-point sampling, the wavefunctions can naturally be grouped according to the $k$-points, which adds an additional layer of parallelization.  




\subsection{Evaluation of $H\Psi$} \label{subsec:hpsi}


The cost of evaluating $H\Psi$ mainly consists of two parts: applying the Fock exchange operator and applying the pseudopotential. 

The algorithm for applying the Fock exchange operator using a planewave discretization is shown in 
Alg.~\ref{alg:Fockalg}. The wavefunctions are distributed according to the band index. For a system with $N_e$ wavefunctions and calculated on $N_p$ processors, each processor holds $N_{e'}=N_e/N_p$ wavefunctions (assuming $N_e$ is divisible by $N_p$). According to Eq.~\eqref{eqn:applyVX}, each wavefunction $\Psi_i$ needs to be multiplied to all other wavefunctions $\{\Psi\}_{j=1}^{N_e}$. This is performed using an \textsf{MPI\_Bcast} operation in line 3 of Alg.~\ref{alg:Fockalg}. Then each processor will solve the $N_{e'}$ Poisson-like equations with FFTs. 
Since each MPI task will eventually receive all $N_e$ wavefunctions, the total communication volume is $N_\text{p}\times N_G\times N_e$ multiplied by the storage cost of a complex number, where $N_G$ is the number of planewaves to store a wavefunction.


\begin{algorithm}
  \caption{Applying the Fock exchange operator in $H\Psi$ }
\begin{flushleft}
   \textbf{INPUT:}  Wavefunctions $\Psi$ distributed according to the band index. \\ 
   \textbf{OUTPUT:} $V_X\Psi$ distributed according to the band index.
\end{flushleft}

\begin{algorithmic}[1]
   \STATE Let $V_X\Psi$ be distributed according to the band index and 
   initialized to zero.
   \FOR {i = 1, $N_e$}
      \IF {the current processor holds $\Psi_i$}
        \STATE Broadcast $\Psi_i$ to all processors
      \ENDIF
      \FOR {j = 1, $N_e$}
        \IF {the current processor holds $\Psi_j$}
        \STATE Solve Poisson-like equation using FFT with respect to the charge-like quantity $\Psi_i^{*}(\vr) \Psi_j(\vr)$, and add the solution to $(V_X\Psi)_j$.
        \ENDIF
      \ENDFOR
   \ENDFOR
\end{algorithmic}
\label{alg:Fockalg}
\end{algorithm}

In order to efficiently carry out Alg.~\ref{alg:Fockalg} on GPUs, we perform a number of optimization steps. 

1. \textit{CUFFT and CUDA custom kernels (band-by-band)}. 
The first step of porting PWDFT onto GPU is to use the CUFFT library and CUDA custom kernels to accelerate the computation of Alg.~\ref{alg:Fockalg}. In this step, all relevant computation( from line 6 to line 10 in Alg.~\ref{alg:Fockalg}) are moved onto GPU in a band-by-band manner. 
In our implementation, the CUDA custom kernels are written to fill the gaps between the CUFFT calls, and there is no CPU-GPU synchronization during the computation. The CPUs are only used for performing MPI communication, and the data copy between CPU and GPU is necessary after the \textsf{MPI\_Bcast} operation. 

2. \textit{Batched implementation}.
Each V100 GPU on the Summit supercomputer has a peak performance of 7.8 TFLOPS and a peak bandwidth of 900 GB/s. The band-by-band implementation above cannot saturate the bandwidth of the GPUs. One way of improving GPU performance is to send more data to the GPU. In the GPU version of PWDFT, instead of sending the data $\Psi_i^{*}(\vr)\Psi_j(\vr)$ one by one, we batch them together and call a batched CUFFT. The corresponding CUDA custom kernels are also changed to a batched fashion. The batched version of code has two benefits: first it increases the computational intensity of the CUDA kernels to take advantage of the computing power of GPU; second, it reduces the latency between CPU and GPU by reducing the number of CUDA kernel launches. 

3. \textit{GPUDirect and CUDA-aware MPI}.
On the Summit supercomputer, the IBM Spectrum CUDA-aware MPI is supported by hardware, which means inter-node GPUs can communicate with each other via MPI without explicitly coping data to the CPU. In PWDFT, we take advantage of this feature, and MPI communication is performed directly on the GPU data. In this step, the wavefunctions are always kept on the GPU, and \textsf{MPI\_Bcast} is performed using the CUDA-aware MPI in a band-by-band manner. This fine-grained (band-by-band \textsf{MPI\_Bcast}) communication creates more opportunity to overlap the GPU computation and the MPI communication, since conceptually the CUDA-aware MPI and GPU computation can be performed simultaneously. 

4. \textit{Single precision MPI}.
As will be seen later in the performance analysis, 
the MPI communication is mainly limited by the bandwidth of network adapters (NIC). To reduce the communication time of the Fock exchange operator applications, we use the single precision format  for sending and receiving the wavefunctions, which reduces the communication volume by half. We also remark that the single precision format is only used in the MPI communication, which means wavefunctions will be converted back to the double precision format for computation. We find that this leads to negligible changes in the accuracy of the rt-TDDFT dynamics. This agrees with observations of ground state DFT calculations~\cite{John-luc2016}. 

\begin{figure}[h]
  \begin{center}{\includegraphics[width=0.48\textwidth]{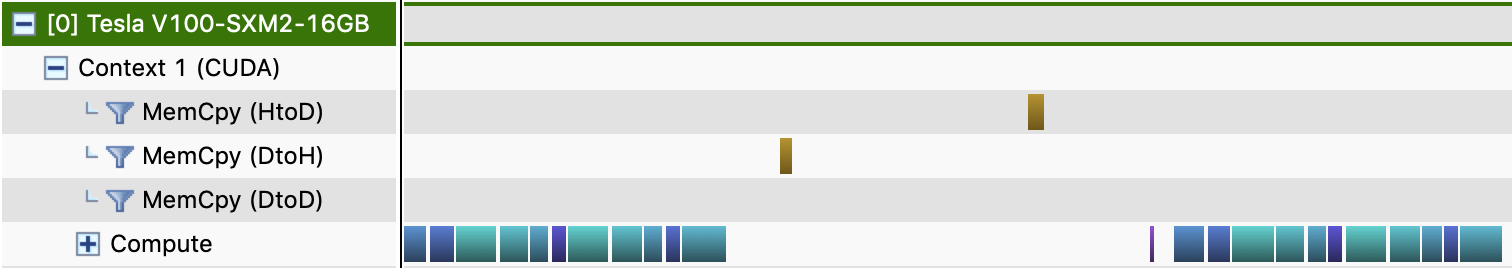}} \end{center}
  \caption{CUDA profiling reveals the implicit CPU-GPU synchronization introduced by the CUDA-aware Spectrum MPI. }
  \label{fig:profile}
\end{figure}

\begin{figure}[h]
  \begin{center}{\includegraphics[width=0.45\textwidth]{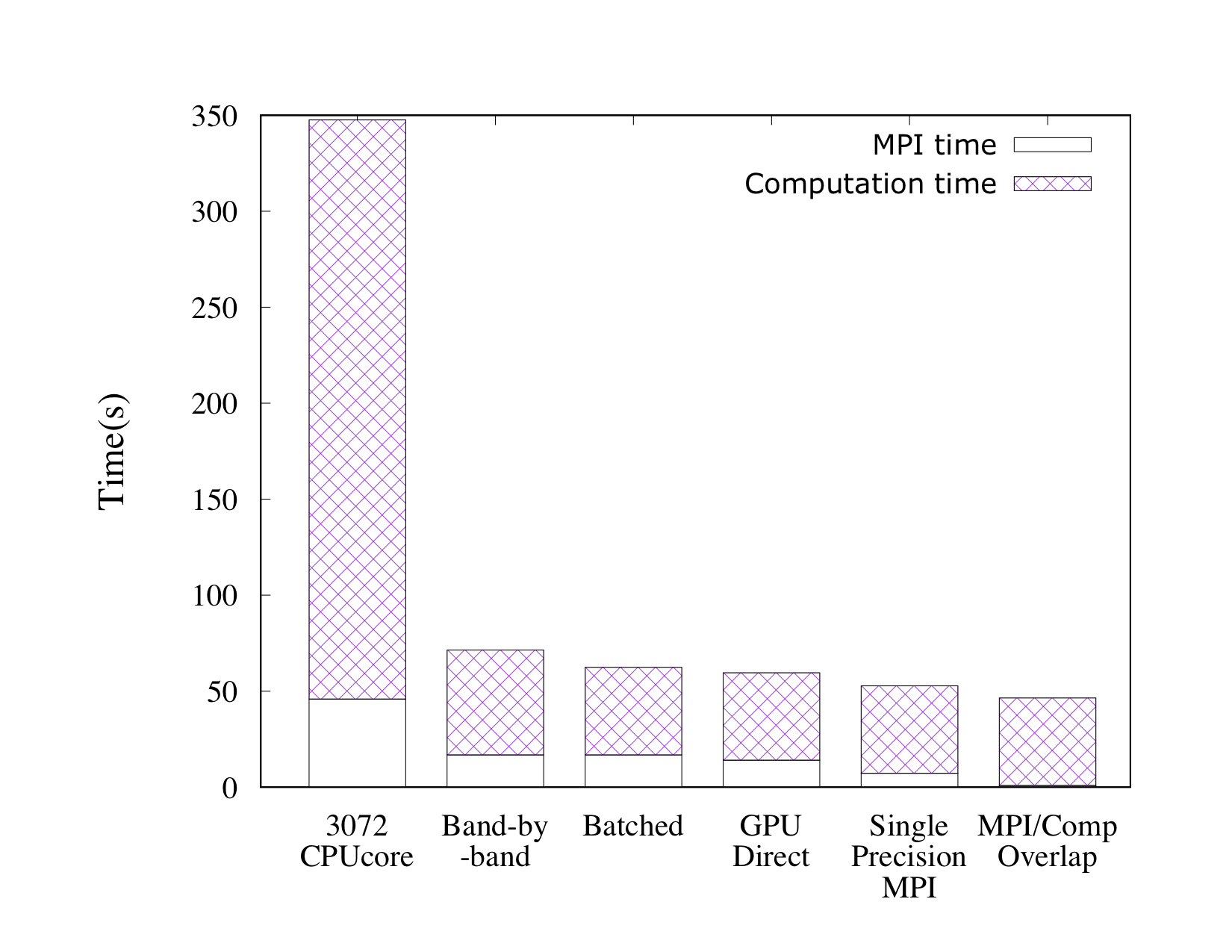}} \end{center}
  \caption{Wall clock time for applying the Fock exchange operator for a system with 1536 silicon 
  atoms. The CPU version uses 3072 CPU cores, and the GPU version uses 72 GPUs.}
  \label{fig:fockspeedup}
\end{figure}

5. \textit{Overlap computation and communication}.
The last step of optimizing the Fock exchange operator calculation is to overlap the MPI communication and computation. Although theoretically CUDA-aware MPI communication overlaps with the GPU computation, we find that the \textsf{MPI\_Bcast} and computation of the Fock exchange operator cannot be fully overlapped when the CUDA-aware \textsf{MPI\_Bcast} is involved on Summit. Profiling of the Fock exchange part shows that there are two synchronized CPU-GPU memory copy operations in the communication step, as shown in Fig.~\ref{fig:profile}. This is caused by the fact that the NIC is connected to the IBM POWER 9 socket as shown in Fig.~\ref{fig:summit}. Thus \textsf{MPI\_Bcast} will first copy the data from GPU to CPU, then the inter-node communication is performed over the NIC. This memory copy operation will introduce the CPU-GPU synchronization, thus the overlapping of computation and communication is disrupted. We get around this issue by performing an asynchronous CPU-GPU memory explicitly, followed by the \textsf{MPI\_Bcast} using CPU instead of the CUDA-aware MPI. We remark that the CUDA-aware MPI are indeed used in other parts of the communication such as \textsf{MPI\_Alltoallv}. Fig.~\ref{fig:fockspeedup} shows that the MPI communication and GPU computation can overlap perfectly, as the MPI communication time is entirely hidden behind the computation time. We stress that the overlapping is achieved based on the fact that CPU and GPUs can work independently, not on the unblocked \textsf{MPI\_Isend}/\textsf{MPI\_Irecv} communication. We have also implemented the round-robin communication strategy~\cite{Ratcliff_2018} via \textsf{MPI\_Send/MPI\_Recv}. We notice that the performance using the round-robin strategy and using \textsf{MPI\_Bcast} is approximately the same on Summit. We also find that the round-robin strategy needs to be carefully implemented to be load-balanced. On the other hand, \textsf{MPI\_Bcast} is a simpler strategy and takes advantage of the fat-tree interconnect topology of Summit.

Fig.~\ref{fig:fockspeedup} shows the reduction of the computational time associated with different stages of optimization. The testing system is a 1536 silicon atoms system, which will be discussed in section~\ref{sec:system}. The CPU version of PWDFT uses 3072 CPU cores (about 73 Summit nodes). The GPU version uses 72 GPUs (12 Summit nodes), and is around 7 times faster than the CPU version in terms of applying the Fock exchange operator.



Besides the application of the Fock exchange operator, we also apply the pseudopotentials to $\Psi$ using the band index parallelization on GPU with CUFFT and CUDA custom kernels. In our implementation, we choose the real space representation for the nonlocal projectors, which can be stored as sparse vectors. This can often be more than $5$ times faster compared to the reciprocal
space implementation when the system size is more than a few hundred atoms~\cite{Wang2001}. In our current implementation, the entire set of local pseudopotentials and nonlocal projectors are stored on every processor. For the largest system tested in this paper with $1536$ silicon atoms, the total memory cost for the nonlocal projectors is approximately 432MB. Each V100 GPU has its 16GB on-chip memory and is therefore sufficient. This simplified implementation allows us to apply the pseudopotentials without any communication cost, and thus fully takes advantage of the computational speed provided by the GPUs.

\subsection{Evaluation of the residuals} \label{subsec:residual}

According to Alg.~\ref{alg:PTCN}, there are two types of residuals to be evaluated in the PT-CN algorithm, denoted by $R_n$ and $R_f$. For simplicity we only discuss the evaluation of $R_f$ (Alg.~\ref{alg:Residualalg}); the evaluation of $R_n$ is similar. 

To calculate the residual $R_f$, the input data $\Psi_f$, $H_f\Psi_f$, and $\Psi_{n+\frac12}$ are first converted from band index parallelization to $G$-space parallelization via an \textsf{MPI\_Alltoallv} operation. Then the local overlap matrix $S$ can be evaluated on GPU by calling the CUBLAS matrix-matrix multiplication routine. Next, the local information from all processors is combined into the global overlap matrix using an \textsf{MPI\_Allreduce} operation, followed by the rotation operation on GPU through a CUBLAS GEMM call. The residual is then calculated by BLAS-1 operations. Finally, the residual $R_f$ will be transposed back to band index parallelization using an \textsf{MPI\_Alltoallv} operation.  In order to reduce the communication cost, the single precision format for $R_f$ is used during the communication step using \textsf{MPI\_Alltoallv}, and is then converted back to the double precision format during the computation step.


\begin{algorithm}
  \caption{ Algorithm of residual calculation.}
\begin{flushleft}
   \textbf{INPUT:}  Wavefunctions $\Psi_f$, $H_f\Psi_f$, $\Psi_{n+\frac{1}{2}}$ distributed according to the band index.\\ 
   \textbf{OUTPUT:} Residual $P_i$ distributed according to the band index.
\end{flushleft}

\begin{algorithmic}[1]
   \STATE Use \textsf{MPI\_Alltoallv} to convert  $\Psi_f$, $H_f\Psi_f$, $\Psi_{n+\frac{1}{2}}$ to the $G$-space parallelization format.
   \STATE Evaluate the local overlap matrix $S_{\text{temp}}=\Psi_f^* H_f\Psi_f$
   \STATE Use \textsf{MPI\_Allreduce} operation on $S_{\text{temp}}$ to obtain the total overlap matrix $S$.
   \STATE Rotate the wavefunctions locally $\Psi_{\text{temp}} = \Psi_f S $
   \STATE Evaluate the residual $R_f = \Psi_f + \frac{\I \Delta t}{2}(H_f \Psi_f-\Psi_\text{temp}) -\Psi_{n+\frac12}$
   \STATE Use \textsf{MPI\_Alltoallv} to convert $R_f$ to the band index parallelization format.
\end{algorithmic}
\label{alg:Residualalg}
\end{algorithm}



%

\subsection{Density Evaluation, Anderson mixing, wavefunction orthogonalization, and others}\label{subsec:densityandothers}

Because the wavefunctions are stored in the band index parallelization format, it is straightforward to evaluate the electron density  $\rho(\vr) = \sum_{i=1}^{N_e} |\psi_i(\vr)|^2$.  This step requires representing the wavefunctions on a real space grid, which can be performed using FFTs, 
followed by an \textsf{MPI\_Allreduce} operation across all MPI tasks. All above calculations are evaluated on the GPU, and the \textsf{MPI\_Allreduce} operation is performed via CUDA-aware MPI. 

The Anderson mixing method for solving the nonlinear equations requires the solution of a least squares problem for each wavefunction~\cite{Anderson1965}. In our calculations, the maximum mixing dimension is set to 20. Thus after evaluating an overlap matrix with respect to the history of wavefunctions $\Psi_f$ and the associated residuals $R_f$, the size of the least squares problem becomes very small, up to $20\times 20$.  The main cost of the Anderson mixing is then due to the evaluation of  overlap matrices that can be performed efficiently using the $G$-space parallelization.

Note that our implementation implies that up to $20$ copies of the wavefunctions are needed for performing the Anderson mixing, which could be expensive if the wavefunctions are all stored on GPUs. However, we may store these wavefunctions on the CPU main memory, which has a large capacity of 512 GB on each computing node of Summit. During the computation, we copy all the wavefunctions corresponding to a single band $i$(up to $20\times N_G$) from CPU to GPU, and the overlap matrices needed for the Anderson mixing can be performed via CUBLAS matrix-matrix multiplications.



At the end of each rt-TDDFT time step, the wavefunctions $\Psi$ will be re-orthogonalized. To improve the parallel efficiency, we again first evaluate an overlap matrix of the type $\Psi^*\Psi$ using the $G$-space parallelization. Then we can perform a Cholesky decomposition on the overlap matrix of size $N_e$, and rotate $\Psi_f$ efficiently due to the $G$-space parallelization.  The Cholesky decomposition is 
calculated on a single GPU with cuSOLVER library, and the subsequent rotation is performed via the GPU \textsf{Trsm} subroutine.

Besides the computationally intensive parts discussed above, 
all other operations such as the evaluation of the Hartree potential, the gradient of the electron density, the local part of the exchange-correlation potential, etc contributes to less than $2\%$ of the computational time on CPUs.
In the GPU version of PWDFT, 
these parts are all parallelized at the CPU level. For example, we parallelize the FFTs associated with the calculation of the gradient of electron density by using distributed FFTW in the Z direction. Such parallelization is important for the overall performance since all other computational intensive parts can be accelerated by up to 40 times on GPUs. We also keep the variables related to the charge density (such as the Hartree potential and the gradient of the electron density) on each MPI task. Hence \textsf{MPI\_AllGatherv} and \textsf{MPI\_Bcast} operation are performed after the computation. 




\section{Setup of the test physical system} \label {sec:system}

We report the efficiency of the GPU version of PWDFT using silicon systems ranging from 48 to 1536 atoms. The supercells are constructed from $1\times1\times3$ to $4\times6\times8$ unit cells, respectively, and each simple cubic unit cell consists of 8 silicon atoms with lattice constant being 5.43 $\angstrom$. In all the tests, the external potential is given by a laser pulse shown in
Fig.~\ref{fig:laser}(a), and its wavelength is $380$nm. The total simulation time is 30 fs, and the time step using the PT-CN method 
is set to 50 as. Thus the total number of rt-TDDFT 
steps is $600$. The stopping criteria is set to $1.0\times10^{-6}$ for the electron density error. 
The average number of SCFs is $22$ and the maximum Anderson mixing dimension 
is set to $20$. We use the SG15 Optimized Norm-Conserving Vanderbilt (ONCV)
pseudopotentials \cite{Hamann2013,SchlipfGygi2015} and HSE06 functionals
\cite{HeydScuseriaErnzerhof2006} in all the following tests. 
The kinetic energy cutoff is set to 
$10$ Hartree. For the system with 1536 atoms, the number of grid points for a wavefunction is $N_G=60\times90\times120=648,000$. This corresponds to a  charge density grid $120\times180\times240$. The Fock exchange operator is evaluated on the wavefunction grid. The number of occupied wavefunctions is 3072. 


\begin{figure}[h]
  \begin{center}
  \subfloat[1536 silicon]{\includegraphics[width=0.20\textwidth]{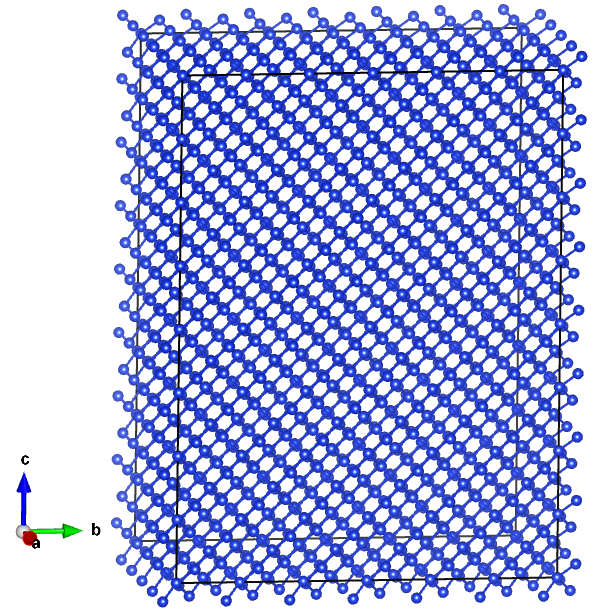}} 
  \subfloat[380nm laser]{\includegraphics[width=0.30\textwidth]{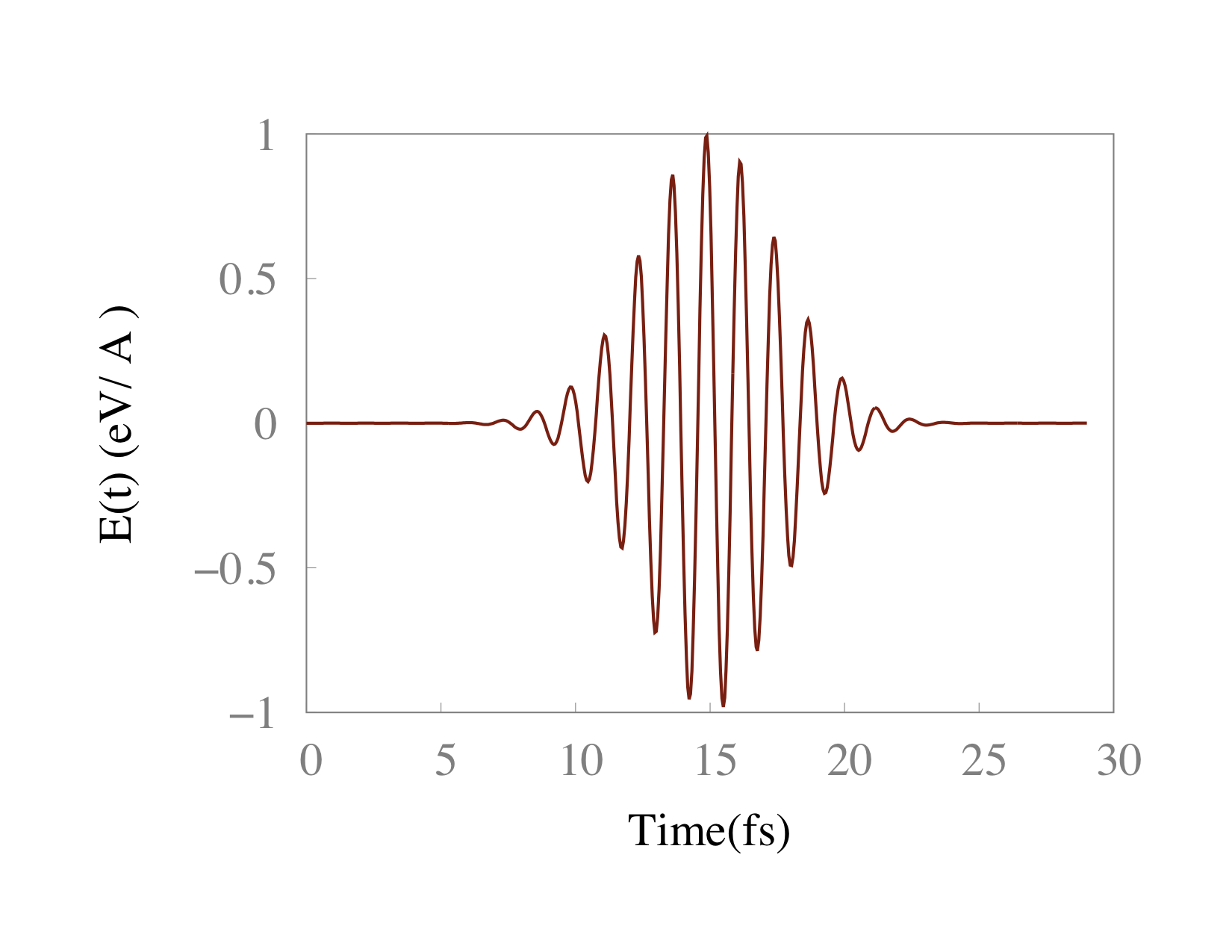}}
  \end{center}
  \caption{Atomic configuration and external laser field.}
  \label{fig:testcase}
\end{figure}


\section{Machine configuration}  \label {sec:machine}

\begin{figure}[h]
  \begin{center}{\includegraphics[width=0.30\textwidth]{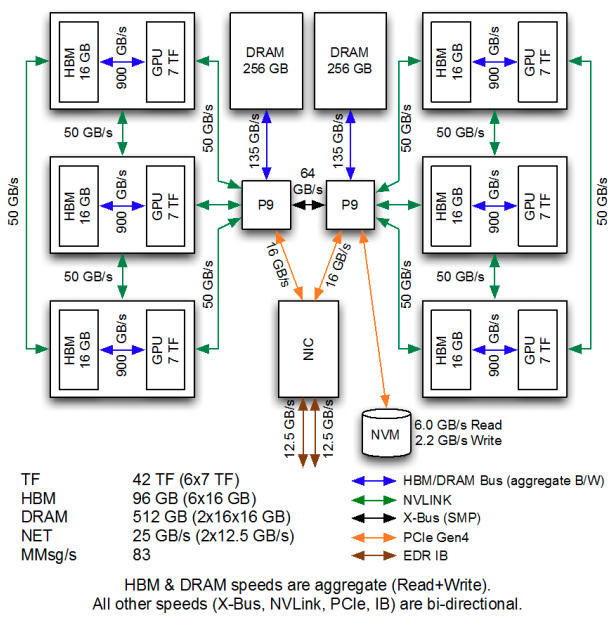}} \end{center}
  \caption{The architecture of a computational node on Summit.}
  \label{fig:summit}
\end{figure}

All numerical tests are performed on the Summit supercomputer. Fig.~\ref{fig:summit} shows the architecture of one of the 4608 
Summit computing nodes. Each computing node consists of two identical groups, and
each group has one IBM POWER 9 
socket and 3 NVIDIA Volta V100 GPUs connected via NVLink with a bandwidth of 50GB/s. 
Each POWER socket has 22 physical CPU cores and share 256GB DDR4 CPU main memory, 
and each V100 GPU has its own 16GB high bandwidth memory. 
The CPU bandwidth is 135GB/s and GPU bandwidth is 900GB/s. 
Each GPU has a theoretical peak performance of 7.8 TFLOPS double precision operations. 
The two groups of hardware are connected via X-Bus with a 64GB/s bandwidth.
The computing nodes are interconnected with a non-blocking fat-tree using a 
dual-rail Mellanox EDR InfiniBand interconnect with a total bandwidth of 25GB/s.


In this paper, we use the MPI+CUDA programming model. In all GPU tests, we use 6 MPI tasks per computing node (3 MPI tasks per socket to fully take
advantage of both CPU-GPU affinity and network adapter), and each MPI task is bound to an
individual GPU. 
For the comparison of the numerical performance, the CPU version of PWDFT only uses the CPU part of the machine. 
In the CPU tests, we use the maximum number of cores allowed by PWDFT. Due to the hybrid parallelization scheme, the maximum number of
CPU cores is the number of wavefunctions. In the case
of 1536 atom silicon system with 3072 wavefunctions, we find that the CPU version of the PWDFT
efficiently scales up to 3072 CPU cores.  


\section{Numerical results} \label{sec:result}

\begin{figure}[h]
  \begin{center}{\includegraphics[width=0.35\textwidth,angle=270]{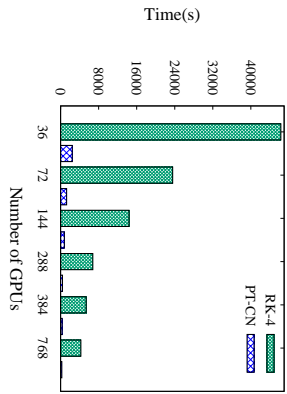}} \end{center}
  \caption{Wall clock time for simulating the 1536 silicon atom system for 50 attoseconds using RK4 and PT-CN methods.}
  \label{fig:ptcn-rk4}
\end{figure}

First, we demonstrate the efficiency of the PT-CN algorithm by comparing it to
the explicit Runge-Kutta 4th order (RK4) integrator for the system with 1536 silicon atoms. The time step for PT-CN is set to 
50 as and the time step for RK4 is 0.5 as. This is close to the largest time step allowed by RK4 due to the stability constraint. 
In Fig.~\ref{fig:ptcn-rk4}, we compare the wall clock time for a 50 as simulation using PT-CN and RK4. 
The PT-CN method can be about 20 times faster compared to the explicit
time integrator RK4 method using 36 GPUs, and becomes 30 times faster when using 
768 GPUs. 
The increase of the speedup factor with respect to the number of GPUs is mainly due to that PT-CN can use a larger step size, and is less impacted by the cost of ``others'' component in section~\ref{subsec:densityandothers} (such as the evaluation of the Hartree potential). A detailed discussion of the scaling of different components will be presented in section \ref{sec:analysis}.


Second, we compare the performance of the GPU version of PWDFT with the CPU one.
One of the most important criteria in high performance computing, especially in the
upcoming exascale era, is power consumption. The power consumption of a single
POWER 9 socket is 190 watt and that of a single NIVIDIA V100 GPU is 300 watt. 
Hence the power cost for each CPU node consisting of 2 POWER 9 CPU sockets is 380 watt and each GPU 
computing node with 6 V100 GPUs and 2 POWER 9 CPU sockets is 2180 watt.
Using 3072 CPU cores (in practice using 73 computing nodes), the total power consumption is $27740$ watt.
The power consumption of 12 GPU nodes is $26160$ watt, which consumes slightly less energy than 73 CPU nodes. 
According to Table \ref{table:total}, in this setup, the GPU version of PWDFT is still 7 times faster compared to the CPU version. 


\begin{figure}[h]
  \begin{center}
  \subfloat[]{\includegraphics[width=0.45\textwidth]{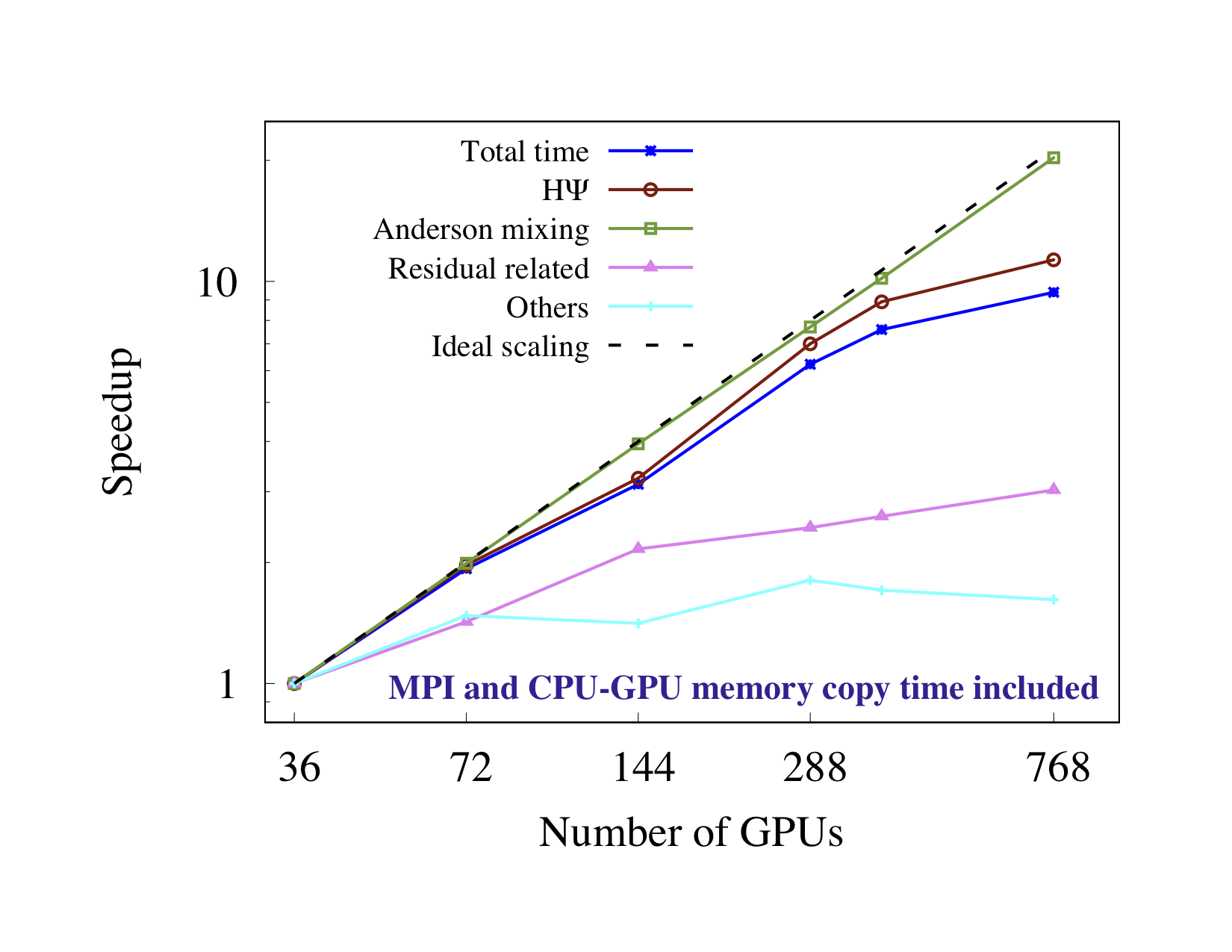}} \\
  \subfloat[]{\includegraphics[width=0.45\textwidth]{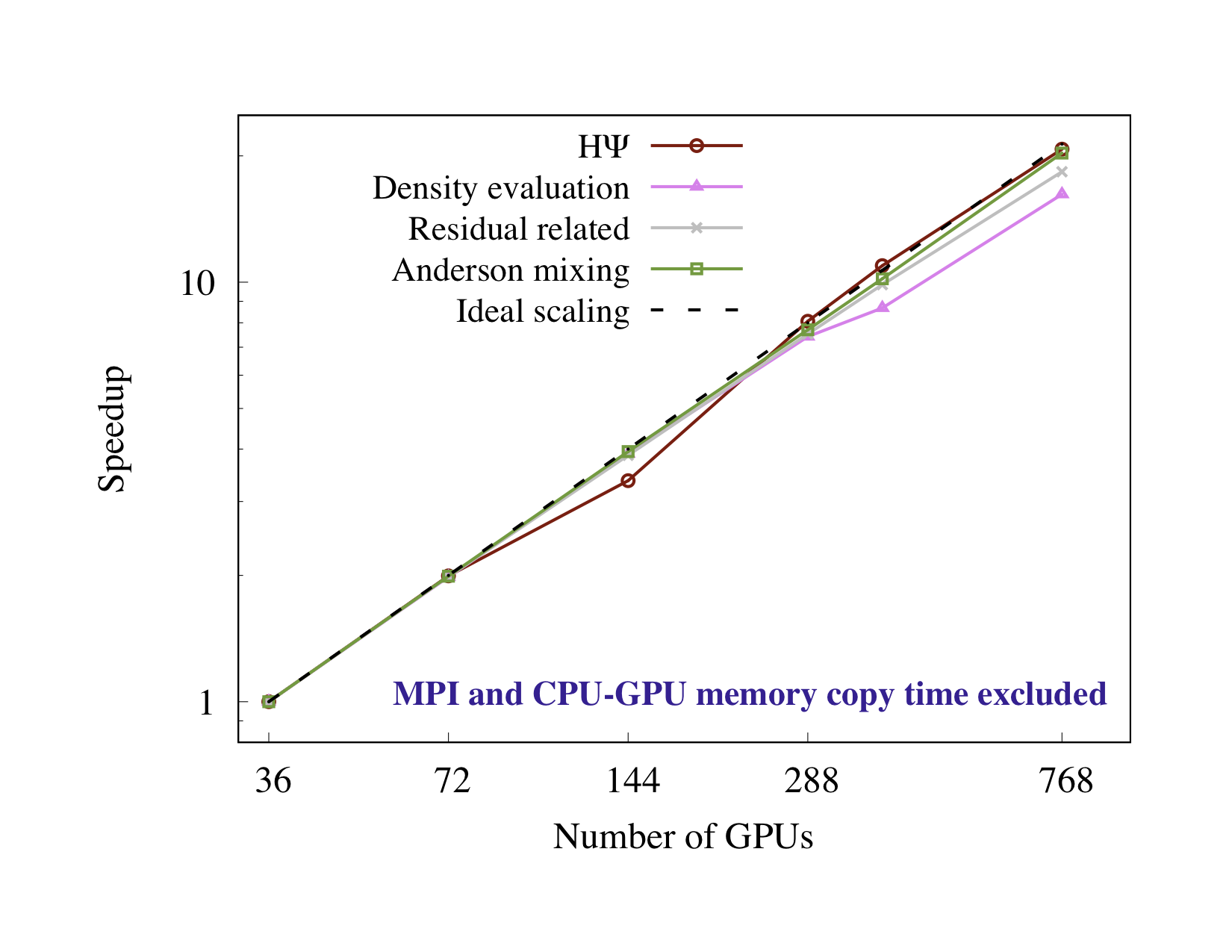}}
  \end{center}
  \caption{Strong scaling: (a) Scaling of the total computation time and different components (MPI and CPU-GPU memory copy time included). (b) Scaling of different components (MPI and CPU-GPU memory time excluded). The wall clock time with 36 GPUs is set as the baseline, which is already 3.7 times faster than the CPU version with 3072 CPU cores.}
  \label{fig:strong_scaling}
\end{figure}





While the CPU version of PWDFT has reached its scaling limit, the GPU version can still scale beyond 72 GPUs. 
Fig.~\ref{fig:strong_scaling}(a) demonstrates the strong scaling of the wall clock time with respect to  
the number of GPUs. We find that our GPU implementation can scale to 768 GPUs, and near ideal scaling is achieved when the GPU number is less than 384. After 768 GPUs, the MPI communication dominates the computational cost, which prevents the code to scale to a larger number of GPUs. According to Table \ref{table:total}, the GPU version is 34 times faster than the CPU version using 3072 CPU cores.  We remark that the strong scaling determines the time to solution, and thus is crucially important in practical calculations. Using the GPU version of PWDFT, we can achieve 260 seconds per TDDFT step (50 attoseconds), which amounts to about 1.5 hours per femtosecond for the 1536 silicon atoms system. 

\begin{figure}[h]
  \begin{center}{\includegraphics[width=0.35\textwidth,angle=270]{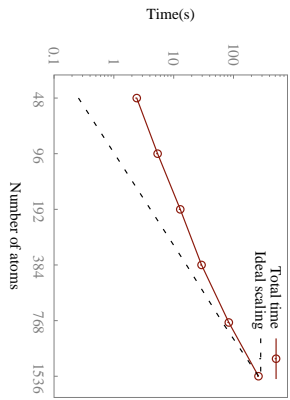}} \end{center}
  \caption{Weak scaling: wall clock time per 50 as for silicon systems with 48 to 1536 atoms.
  The number of GPUs used are always set to half of number of atoms in the calculation.  The ``ideal scaling'' here scales as $O(N_{atom}^2)$.
  }
  \label{fig:weakscaling}
\end{figure}

The weak scaling using PT-CN method for simulating systems consisting of 48 to 1536 atoms for 50 as is shown in Fig.~\ref{fig:weakscaling}.  The number of GPUs is set to $\frac{1}{2}N_\textsf{atom}$. The GPU version of PWDFT exhibits excellent weak scaling property. Since the computational complexity of hybrid functional rt-TDDFT simulation scales as $O(N_{atom}^3 \log N_{atom})$, the ideal scaling should be $O(N_{atom}^2)$ as in Fig.~\ref{fig:weakscaling}, neglecting the logarithmic factor. We find that for small systems, thanks to GPU acceleration, the Fock exchange operator applications has not yet dominated the computational cost, and hence our implementation scales even better than that indicated by the ideal scaling. Even with the system size increases to $1536$ atoms, the weak scaling is still very close to the ideal scaling. 

Here for a smaller system with 192 atoms, the simulation of 50 as with 96 GPUs is only 16 seconds. This means that each femtosecond simulation takes around $5$ minutes. Even a picosecond rt-TDDFT simulation with hybrid functionals is now within reach and would take approximately $4$ days. 


\section{Performance Analysis} \label{sec:analysis}

In this section, we present a more detailed analysis of the performance of our GPU implementation. 

First, we discuss the memory usage of the PT-CN method. In the GPU version of PWDFT, the most memory demanding part is the Anderson mixing, which requires up to 20 copies of wavefunction $\Psi$. For the system with 1536 atoms, each wavefunction takes 10MB ($N_G=648,000$ multiplied by the cost of a complex number in double precision format). Using 36 GPUs, each MPI holds less than 100 wavefunctions (1GB). Then Anderson mixing requires less than 20 GB memory per MPI. There are 6 MPIs per computing node, and the total usage of the CPU main memory per node is less than 120GB, which is smaller than the limit of a node on Summit (512GB). Hence our implementation of PT-CN effectively takes advantage of the fat node configuration.  

The total number of double precision floating point operations (FLOP) for the 1536 silicon system per TDDFT step 
is $3.87\times10^{16}$. This is collected via the CUDA profiling tool NVPROF. Although NVPROF only collects the total number of FLOP on the GPUs, in our implementation the CPU is only responsible for computing quantities labeled as ``others'' as in section~\ref{subsec:densityandothers}. The floating point operations per second (FLOPS) is then calculated as $\frac{\textsf{total FLOP}}{\textsf{(number of GPUs)} \times \textsf{(total time)}}$, and the corresponding FLOPS efficiency is $\frac{ \textsf{FLOPS}} {\textsf{7.8 TFLOPS}}$. The efficiency of GPU version of PWDFT 
is $5.5\%$ when using 36 GPUs, and goes down to $2\%$ using 768 GPUs. The low FLOPS efficiency of GPU version of PWDFT is mainly caused by the fact that most FLOP is contributed by the FFTs in Fock exchange operator calculation. The FFT operations on GPU is mainly limited by the CPU-GPU bandwidth rather than the computational kernel. For instance, we find that CUFFT execution reaches about $11\%$ of the peak performance of the V100 GPU in our implementation, and the result is comparable to the performance of CUFFT reported by NVIDIA ~\cite{NVIDIAReport2016}. The above analysis can be supported by evaluating the average required bandwidth of CUFFT and CUDA custom kernels during the Fock exchange operator calculation. We find that the GPU version of PWDFT achieves approximately $90\%$ of the GPU maximum bandwidth in all tests. Such high GPU memory bandwidth utilization indicates that our calculation is mainly bounded by the hardware memory bandwidth, rather than the FLOPS.

\begin{figure}[h]
  \begin{center}{\includegraphics[width=0.35\textwidth,angle=270]{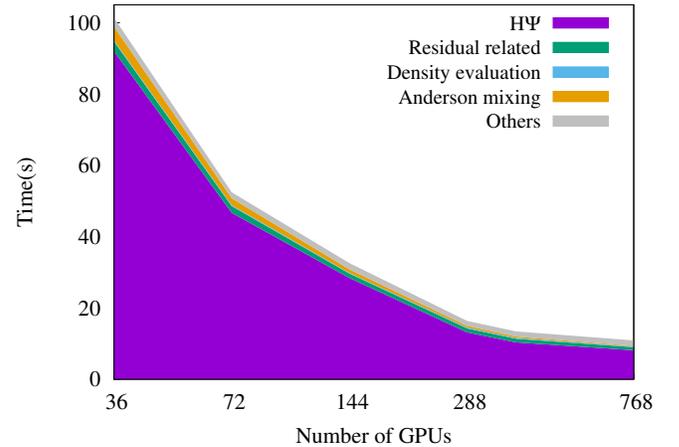}} \end{center}
  \caption{ The total time of a single SCF step and the contributions of each part using different number of GPUs. 
  }
  \label{fig:parts_time}
\end{figure}

The total time of a single SCF step can be divided into 5 parts:  $H\Psi$, 
 Anderson mixing, the residual related part, the electron density evaluation, and others. The contributions of each part to the total time is listed in Table~\ref{table:total} and shown in Fig.~\ref{fig:parts_time}.  The scaling of different computational components of the GPU version of PWDFT is shown in Fig.~\ref{fig:strong_scaling}(a). Since the total computation time is dominated by the application of the Fock exchange operator, the scaling of $H\Psi$ is similar to that of the total time. In each TDDFT step, 24 Fock exchange operator evaluation is performed (22 in the SCF calculation, one in total energy evaluation, and one before SCF calculation to evaluate the residual term $R_n$), and this contributes to $93\%$ of the total FLOP.   
The Anderson mixing step scales well with respect to the number of GPUs. The residual related part also scales with the number of GPUs. Its scaling is mainly limited by the \textsf{MPI\_Alltoallv} and \textsf{MPI\_Allreduce} operations. 
The ``others'' components as in section~\ref{subsec:densityandothers} are all parallelized on CPUs. As shown in Fig.~\ref{fig:parts_time}, ``others'' does not scale with the number of GPUs. It contributes $2.6\%$ of the total time of a single SCF when using 36 GPUs, and grows up to $15\%$ when using 768 GPUs. Such scaling behavior is mainly caused by the fact that ``others'' is dominated by the MPI communication of the density related variables using \textsf{MPI\_Bcast}.

\begin{table*}[ht]
\caption{
Wall clock time of the computationally intensive components for calculating a 1536 silicon atom system. 
The speedup factor is based on the best CPU implementation with 3072 CPU cores using about 73 computing nodes, and the wall clock time is 8874s. 
}
\centering
\begin{tabular}{|p{0.28\linewidth} p{0.06\linewidth}p{0.06\linewidth} p{0.06\linewidth}
p{0.06\linewidth}p{0.06\linewidth} p{0.06\linewidth}p{0.06\linewidth}p{0.06\linewidth} | }
\hline
\textbf{Number of GPUs} & 36  & 72  & 144 & 288 & 384 & 768 &  1536 & 3072 \\
\hline
Fock exchange operator MPI         &0.71&0.89&1.25&1.83&1.99&3.72&6.06&8.074 \\
Fock exchange operator computaion  &90.99&45.61&27.05&11.27&8.31&4.38&2.44&1.43 \\
Fock exchange operator total time  &91.7&46.5&28.3&13.1&10.3&8.1&8.5&9.5 \\
Local and semi-local part          & 0.337 & 0.169 & 0.087& 0.043& 0.0316& 0.0158& 0.00805& 0.00404\\
$H\Psi$ total time & 92.04&46.67&28.39&13.14&10.33&8.12&8.51&9.50 \\

\hline
Wavefunction MPI\_Alltoallv  & 0.884& 0.561& 0.313& 0.227& 0.212& 0.280& 0.095& 0.056\\
$\left\langle\Psi|\Psi\right\rangle$ MPI\_Allreduce & 0.354& 0.593& 0.552& 0.676& 0.667& 0.523& 0.522& 0.5243\\
Computation    & 1.43 & 0.72 & 0.37 & 0.19 & 0.145 & 0.078 & 0.04 & 0.023\\
Residual related total time & 2.67& 1.87& 1.24& 1.09& 1.02& 0.88& 0.66& 0.60\\
\hline
CPU-GPU memory copy & 1.64235 & 0.8004& 0.4094& 0.2018& 0.1477& 0.0746& 0.0395& 0.0202\\
Computation  time   & 2.3     & 1.16     & 0.59 & 0.31 & 0.265 & 0.142 & 0.073 & 0.04\\
Anderson mixing total time& 3.94    & 1.98     & 1.00 & 0.51 & 0.387 & 0.194 & 0.102 & 0.0553\\
\hline
Computation  time        & 0.1349 & 0.0672 & 0.0341 & 0.0170 & 0.0124 & 0.0062 & 0.0032 &  0.0016 \\
MPI\_Allreduce            & 0.123 & 0.176 & 0.152 & 0.224 & 0.219 & 0.160 & 0.164 & 0.171 \\
Density evaluation total time & 0.258  & 0.243 & 0.186 & 0.241 & 0.232 & 0.167 & 0.167 & 0.172 \\
\hline
Others               & 2.66 & 1.98 & 1.72 & 1.54 & 1.57 & 1.73 & 1.66 & 1.85 \\
\textbf{per SCF time} &101.36 &52.4&32.5&16.4&13.4&10.9&10.9&12.1\\
\hline

\textbf{Total time} &2453.8&1269.1&783.0&393.9&323.2&260.9&262.5&286.6\\
\textbf{Total speedup}    &3.6x&7.0x&11.3x&22.5x&27.4x&34x&33.8x&30.9x\\
\textbf{$H\Psi$ percentage}    &90\%& 88.3\%& 87\%& 80\%& 76.7\%& 74.6\%& 77.8\%& 79.6\% \\
\hline
\end{tabular}
  \label{table:total}
\end{table*}

\begin{table*}[ht]
\caption{Breakdown of the total time into the time for MPI, CPU-GPU memory copy and computation.  The CPU-GPU memory copy time and MPI time are all gathered in the runtime phase, the computational time is calculated by removing all the communication time from the total time in Table \ref{table:total}. }
\centering
\begin{tabular}{|p{0.25\linewidth}  p{0.06\linewidth}p{0.06\linewidth} p{0.06\linewidth}
p{0.06\linewidth}p{0.06\linewidth}p{0.06\linewidth}p{0.06\linewidth}p{0.06\linewidth} |p{0.06\linewidth} }
\hline
\textbf{Number of GPUs} & 36  & 72  & 144 & 288 & 384 & 768 &  1536 & 3072 \\
\hline
CPU-GPU memory copy time &60.80&29.94&16.04&8.57&6.79&4.15&2.82&2.24\\
\hline
MPI\_Alltoallv time  &20.97&13.34&7.40&5.38&4.99&6.64 &2.41&0.68\\
MPI\_Allreduce time &11.50&18.39&16.70&21.27&21.15  &16.19&16.44&16.62\\
MPI\_Bcast time     &18.78&20.89&31.06&44.54&48.13&92.26&146.15&193.89\\
MPI\_AllGatherv time    &0.44&1.12&1.30&1.35&1.52&1.38&0.98&1.24\\
\textbf{MPI total time} &51.69&53.74&56.45&72.54&75.79&116.47&165.97&212.43\\
\hline
\textbf{Computational time} &2341.40&1185.42&710.54&312.83&240.60&140.34&93.73&71.96\\

\hline
\end{tabular}
  \label{table:part}
\end{table*}

Next, we discuss the scalability of different computational components in Fig.~\ref{fig:strong_scaling}(b). Note that no MPI communication or CPU-GPU memory copy time is included in this figure. We find that nearly all computational time scale well with respect to the number of GPUs. The only computational part that does not scale is the Cholesky decomposition used in the orthogonalization and it is not shown in Fig.~\ref{fig:strong_scaling}(b). This part only takes 0.017s on GPU for the 1536 atom system  and is calculated once every TDDFT step. Thus it is negligible in the rt-TDDFT calculation. The scaling of the computational time clearly shows that the scaling bottleneck is not the computation. Therefore the main bottleneck comes from the data moment operations, which include both the CPU-GPU memory copy and the MPI communication. 

\begin{figure}[h]
  \begin{center}
    {\includegraphics[width=0.35\textwidth,angle=270]{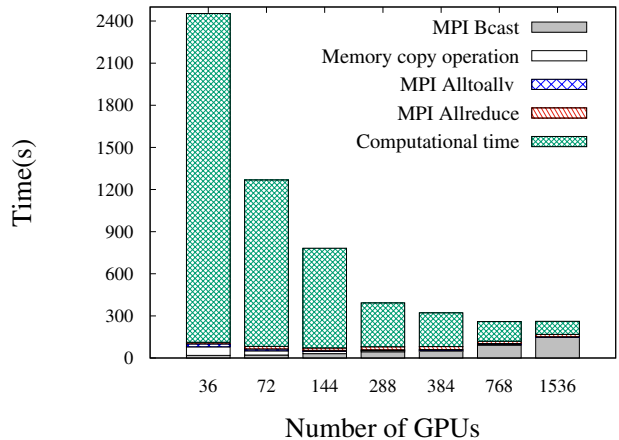}} 
  \end{center}
  \caption{Strong scaling of different operations: MPI communication, CPU-GPU memory 
  copy operation and computation.}
  \label{fig:mpi_mem_com}
\end{figure}

The breakdown of the wall clock time in terms of the MPI communication, CPU-GPU memory copy and computation shown in Fig. ~\ref{fig:mpi_mem_com}. The detailed numbers are reported in Table \ref{table:part}. Since memory copies are mostly performed over wavefunctions within each node, the CPU-GPU memory copy operations scale well with respect to the number of GPUs. The \textsf{MPI\_Alltoallv} operations is mainly used in the hybrid parallelization scheme to convert the distribution formats of wavefunctions, and is also found to be scalable. The \textsf{MPI\_AllGatherv} operationis performed after the exchange-correlation potential is calculated via \textsf{Libxc}~\cite{LEHTOLA2018}. It contributes less than $0.6\%$ of total time and is thus negligible. The \textsf{MPI\_Bcast} operation is mainly used in the Fock exchange operator to broadcast one wavefunction to all GPUs. The \textsf{MPI\_Allreduce} operation is performed for computing the charge density and to compute the overlap matrix. These two components are the communication bottleneck.  We notice that in our testing results, there are some fluctuations in terms of the communication time in Table ~\ref{table:part}. For example, the time for \textsf{MPI\_Allreduce} peaks at around 288 and 384 GPUs, and the time for \textsf{MPI\_Alltoallv} has a local peak at 768 GPUs. We confirm that such fluctuation can be repeatedly observed on Summit with the same configuration, and would like to investigate the origin of such fluctuation in the future. 

Let us now analyze the performance of \textsf{MPI\_Bcast} from the receiving side.  In the Fock exchange operator calculation, each node receives 3072 wavefunctions. Each wavefunction consists of $N_G=648,000$ complex numbers, which is 5.0MB in the single precision format.  Thus total communication volume is $3072\times5.0MB=15.36$GB . The communication time without overlapping with computation is about 7 seconds with 768 GPUs, thus the MPI communication speed is $15.36$GB/$7s=$ 2.2 GB/s. The Summit supercomputer has two NICs connecting to two POWER 9 sockets, respectively. The communication bandwidth for each NIC is 12.5 GB/s. Since we have three MPI tasks per socket,  the network bandwidth utilization rate is about $52.7\%$ ($3\times 2.2/12.5$) from the receiving side. In the GPU version of PWDFT, the CPU \textsf{MPI\_Bcast} operation is overlapped with the GPU computation, and the MPI time shown at Table ~\ref{table:total} is part of the total communication time. For example, in the 768 GPU case, almost half of the MPI communication time is overlapped by the computation time. 
Besides the wavefunction \textsf{MPI\_Bcast}, we also have the \textsf{MPI\_Bcast} of the gradient of charge density, etc. Data size of the charge density is 40MB and it is also network bandwidth limited.  
The \textsf{MPI\_Allreduce} operation is mainly used to evaluate the overlap matrix and charge density. Both operations are performed around 24 times in a single TDDFT step. The data size of the overlap matrix and the charge density vector are 144 MB and 40 MB, respectively. Hence the total data size for \textsf{MPI\_Allreduce} is 4.4 GB per time step. This is less than the communication cost of \textsf{MPI\_Bcast} but is of the same order of magnitude.

\section{Conclusion} \label{sec:conclusion}

In this paper, we presented the GPU version of PWDFT for performing rt-TDDFT calculations with hybrid exchange-correlation functional.  Our implementation is based on the planewave discretization and the parallel transport (PT) formulation. The PT formulation is used to increase the size of the time step, and therefore reduces the frequency of the Fock exchange operator applications. The multi-GPU implementation reduces the time for applying the Fock exchange operator, as well as other computationally intensive components in rt-TDDFT calculations. This is achieved by carefully rewriting all computational intensive parts with CUDA, and by other techniques to reduce the MPI communication time in the heterogeneous environment. The performance is demonstrated on the Summit supercomputer, and the main techniques can be transferred to other rt-TDDFT as well as ground state DFT software packages.  We also found that on the scalability is mainly limited by the network bandwidth on the Summit supercomputer. Hence we expect the parallel performance could scale further with improved network bandwidth on future supercomputers.

\begin{acks}
This work was partially supported by the National Science Foundation
under Grant No. 1450372, No. DMS-1652330  (W. J. and L. L.), and by the
Department of Energy under Grant No. DE-SC0017867 (L. L.), 
and by the Department of Energy Theory of
Materials (KC2301) program under Contract No. DE-AC02-05CH11231 (L. W.). 
We thank the Oak Ridge National Labratory 
leadship computing facility for making the computational resources
available. We thank Zhanghui Chen and Mauro Del Ben for helpful discussions.
\end{acks}

%
\bibliographystyle{ACM-Reference-Format}
\bibliography{ptref}

%

\end{document}